\documentclass[aps,pra,twocolumn]{revtex4}%
\usepackage{graphics}
\usepackage{amsmath}
\usepackage{graphicx}
\usepackage{amsfonts}
\usepackage{amssymb}
\usepackage{float}
\usepackage{longtable}
\usepackage{epsfig}
\usepackage{latexsym}
\usepackage{theorem}
\usepackage{bbm}
\usepackage{bm}
\usepackage{verbatim}
\usepackage{psfrag}%
\DeclareMathOperator*{\Tr}{Tr}
\newtheorem{theorem}{Theorem}

\newtheorem{remark}[theorem]{Remark}

\begin{document}
\title{Finite-resource teleportation stretching for continuous-variable systems}
\author{Riccardo Laurenza}
\affiliation{Computer Science and York Centre for Quantum Technologies, University of York,
York YO10 5GH, United Kingdom}
\author{Samuel L. Braunstein}
\affiliation{Computer Science and York Centre for Quantum Technologies, University of York,
York YO10 5GH, United Kingdom}
\author{Stefano Pirandola}
\affiliation{Computer Science and York Centre for Quantum Technologies, University of York,
York YO10 5GH, United Kingdom}

\begin{abstract}
We show how adaptive protocols of quantum and private communication through
bosonic Gaussian channels can be simplified into much easier block versions
that involve resource states with finite energy. This is achieved by combining
the adaptive-to-block reduction technique devised earlier [S. Pirandola
\textit{et al.}, Nat. Commun. \textbf{8}, 15043 (2017)], based on
teleportation stretching and relative entropy of entanglement,~with an
alternative simulation of Gaussian channels recently introduced by
Liuzzo-Scorpo \textit{et al.} [Phys. Rev. Lett. \textbf{119}, 120503 (2017)].
In this way, we derive weak converse upper bounds for the secret-key capacity
of phase-insensitive Gaussian channels, which approximate the optimal limit
for infinite energy. Our results apply to both point-to-point and
repeater-assisted private communications.

\end{abstract}
\maketitle

\section{Introduction}

Establishing the ultimate limits of quantum and private communications is
important~\cite{kimbleQnet,pirsQnet}, not only to explore the boundary of
quantum mechanics but also to provide benchmarks for testing the practical
performance of experimental and technological implementations. This problem is
important for quantum systems of any dimension~\cite{NiCh,QIbook,Watrous} and,
in particular, for infinite-dimensional ones, also known as
continuous-variable (CV) systems~\cite{SamRMPm,RMP,Alex,Gerry}. In quantum
information and quantum optics, the most important CV\ systems are the bosonic
modes of the electromagnetic field~\cite{SamRMPm}, which are typically used at
the optical or telecom wavelengths. In any protocol of quantum communication,
such modes are subject to loss and noise, and the most typical and basic model
for such kind of decoherence is the single-mode Gaussian channel.

It is known that protocols of private communication and quantum key
distribution (QKD) are limited in both rate and distance due to decoherence,
no matter if the communication line is a free-space link or a fiber
connection. This limitation is perhaps best simplified by the rate-loss
scaling of ideal single-photon BB84 protocol~\cite{BB84} whose optimal rate
scales as $\eta/2$ secret bits per channel use, where $\eta$ is the
transmissivity of the channel. Recently, this fundamental rate-loss limit has
been fully characterized. By optimizing over the most general key-generation
protocols, in Ref.~\cite{PLOB} we established the secret-key capacity of the
lossy channel to be $K(\eta)=-\log_{2}(1-\eta)$, which is about $1.44\eta$
secret bits per channel use at long distances ($\eta\simeq0$). This result
sets a general benchmark for quantum
repeaters~\cite{Rep1,Rep2,Rep3,Rep4,Rep5,Rep6,Rep7,Rep8,Rep9,Rep10,Rep12,Rep13,Rep14,Rep15,Rep16,Rep18,Rep13bis,bench5,LoPiparo1,LoPiparo1bis,LoPiparo2,MihirRouting}
and completes a long-standing investigation started back in
2009~\cite{RevCohINFO,ReverseCAP}, when the best known lower bound was discovered.

The main technique that led to establishing the previous capacity is based on
a suitable combination of two ingredients, the relative entropy of
entanglement (REE)~\cite{RMPrelent,VedFORMm,Pleniom} suitably extended from
states to channels (using results from
Refs.~\cite{Donaldmain,Matthias1a,Matthias2a}), and teleportation stretching,
which reduces any adaptive (feedback-assisted) quantum protocol over an
arbitrary channel into a much simpler block version. This latter technique is
a full extension and generalization of previous
approaches~\cite{B2main,MHthesis,Niset,Wolfnotes} that only worked for
specific classes of channels and were designed to reduce quantum error
correcting code protocols into entanglement distillation. Without doubts, the
generalization to an arbitrary task over an arbitrary quantum channel has been
one of the key insights of Ref.~\cite{PLOB}, and this has been widely
exploited in recent literature, with a number of follow-up papers in the area
of quantum Shannon theory~\cite{QIbook}, e.g., on strong converse rates,
broadcast capacities, etc. See Ref.~\cite{TQCreview}\ for a recent review on
these topics and Refs.~\cite{TQCreview,PLB} for rigorous proofs of some
related claims.

The core of teleportation stretching is the idea of channel simulation, where
an arbitrary quantum channel is replaced by local operations and classical
communication (LOCC) applied to the input and a suitable resource
state~\cite{PLOB}. This powerful idea is rooted in the protocol of
teleportation~\cite{tele,telereview} and first proposed in Ref.~\cite{B2main},
despite originally limited to the simulation of Pauli
channels~\cite{SougatoBowen} (see also Ref.~\cite{nonPauli}). Later, this core
idea was extended to generalized teleportation
protocols~\cite{MHthesis,Leung,Wolfnotes} and CV teleportation~\cite{teleCV}
in Refs.~\cite{Niset,CiracCV}. The final and more general form involves a
simulation via arbitrary LOCCs, as formulated in Ref.~\cite{PLOB}. In
particular, the simulation of bosonic channels is typically asymptotic, which
means that they need a suitable limit over sequences of resource states, which
comes from the fact that the Choi matrices of such channels are asymptotic
states~\cite{PLOB}. Most importantly, such a simulation needs a careful
control of the simulation error as first considered in Ref.~\cite{PLOB},
otherwise technical divergences may appear in the results. This crucial aspect
is discussed in detail in Ref.~\cite{TQCreview}, which also provides a direct
comparison of the various simulation techniques appeared in the literature.

Here we consider a different type of simulation for bosonic Gaussian channels,
which is based on finite-energy two-mode Gaussian states as recently
introduced in Ref.~\cite{GerLimited}. We use this particular simulation at the
core of teleportation stretching in order to simplify adaptive protocols. This
not only represents an interesting design (with potential applications beyond
this work) but also allows us to derive upper bounds for the secret-key
capacities of phase-insensitive Gaussian channels which approximate well the
asymptotic results of Ref.~\cite{PLOB}.

The paper is organized as follows. In Sec.~\ref{sec1}, we review the tool of
channel simulation. In Sec.~\ref{sec2} we use this tool with teleportation
stretching, deriving a single-letter bound for single-mode Gaussian channels.
This bound is explicitly computed in Sec.~\ref{sec3}, where it is also
compared with the infinite-energy one of Ref.~\cite{PLOB}. Theory is then
extended to a chain of quantum repeaters in Sec.~\ref{sec4}. Finally,
Sec.~\ref{sec5} is for conclusions.

\section{Simulation of bosonic channels\label{sec1}}

\subsection{Preliminaries}

As discussed in Ref.~\cite{PLOB} an arbitrary quantum channel $\mathcal{E}$
can be simulated by a trace-preserving LOCC $\mathcal{T}$ and a suitable
resource state $\sigma$, i.e.
\begin{equation}
\mathcal{E}(\rho)=\mathcal{T}(\rho\otimes\sigma)~. \label{locc}%
\end{equation}
A channel is called $\sigma$-stretchable if it has $\sigma$ as a resource
state\ via some LOCC simulation as in Eq.~(\ref{locc}). An important case is
when the channel is Choi-stretchable, which means that the resource state can
be chosen to be its Choi matrix $\sigma=\rho_{\mathcal{E}}:=\mathcal{I}%
\otimes\mathcal{E}(\Phi)$, with $\Phi$ being a maximally entangled state. For
a bosonic channel, the maximally entangled state is an Einstein-Podolsky-Rosen
(EPR) state with infinite energy, so that the Choi matrix of a bosonic channel
is energy-unbounded. For this reason one has to work with a sequence of
two-mode squeezed vacuum states~\cite{RMP} $\Phi^{\mu}$ with variance
$\mu=\bar{n}+1/2$, where $\bar{n}$ is the average number of thermal photons in
each mode. By definition, the EPR state is defined as $\Phi:=\lim_{\mu}%
\Phi^{\mu}$ and the Choi matrix of a bosonic channel $\mathcal{E}$ is defined
by
\begin{equation}
\rho_{\mathcal{E}}:=\lim_{\mu}\rho_{\mathcal{E}}^{\mu},~~\rho{_{\mathcal{E}%
}^{\mu}}=\mathcal{I}\otimes\mathcal{E}(\Phi^{\mu})~. \label{seqq}%
\end{equation}
This means that the simulation needs to be asymptotic, i.e., of the
type~\cite{NoteBELL}
\begin{equation}
\mathcal{E}(\rho)=\lim_{\mu}\mathcal{T}(\rho\otimes\rho{_{\mathcal{E}}^{\mu}%
)}~. \label{asyCC}%
\end{equation}

In Ref.~\cite{PLOB}, we identified a simple sufficient condition for a quantum
channel to be Choi-stretchable, even asymptotically as in Eq.~(\ref{asyCC}%
):\ teleportation covariance. In the bosonic case, a channel $\mathcal{E}$ is
teleportation-covariant if, for any random displacement $D$ (as induced by CV
teleportation~\cite{teleCV,telereview}), we may write
\begin{equation}
\mathcal{E}(D\rho D^{\dagger})=V\mathcal{E}(\rho)V^{\dagger},
\end{equation}
for some unitary $V$. It is clear that bosonic Gaussian channels are
teleportation covariant and, therefore, Choi-stretchable, with asymptotic
simulation as in Eq.~(\ref{asyCC}).

\subsection{Simulation of Gaussian channels with finite-energy resource
states}

Recently, Ref.~\cite{GerLimited} has shown that all single-mode
phase-insensitive Gaussian channels can be simulated by applying CV
teleportation to a particular class of Gaussian states as the resource.
Consider a single-mode Gaussian state with mean value $\bar{x}$ and covariance
matrix (CM) $\mathbf{V}$~\cite{RMP}. The action of a single-mode Gaussian
channel can be expressed in terms of the statistical moments as
\begin{equation}
\bar{x}\rightarrow\mathbf{T}\bar{x},~~\mathbf{V}\rightarrow\mathbf{TV}%
\mathbf{T}^{T}+\mathbf{N,} \label{Act}%
\end{equation}
where $\mathbf{T}$ and $\mathbf{N}=\mathbf{N}^{T}$ are $2\times2$ real
matrices satisfying suitable conditions~\cite{RMP}. In particular, the channel
is called {phase-insensitive if these two matrices take the specific diagonal
forms}
\begin{equation}
\mathbf{T}=\sqrt{\eta}\mathbf{I},~~\mathbf{N}=\nu\mathbf{I} \label{1mode}%
\end{equation}
where $\eta\in\mathbb{R}$ is a transmissivity parameter, while $\nu\geq
0$\ represents added noise.

According to Ref.~\cite{GerLimited}, a phase-insensitive Gaussian channel
$\mathcal{E}_{\eta,\nu}$ can be simulated as follows~\cite{NoteBELL2}
\begin{equation}
\mathcal{E}_{\eta,\nu}(\rho)=\mathcal{T}_{\eta}(\rho\otimes\sigma_{\nu}),
\label{Finn}%
\end{equation}
where $\mathcal{T}_{\eta}$ is the Braunstein-Kimble protocol with gain
$\sqrt{\eta}$~\cite{teleCV,teleMANCIO}, and $\sigma_{\nu}$ is a zero-mean
two-mode Gaussian state with CM~\cite{Notation}%
\begin{equation}
\mathbf{V}(\sigma_{\nu})=\left(
\begin{array}
[c]{cc}%
a\mathbf{I} & c\mathbf{Z}\\
c\mathbf{Z} & b\mathbf{I}%
\end{array}
\right)  , \label{resState}%
\end{equation}
where~\cite{GerLimited}
\begin{align}
a  &  =\frac{2b+(\eta-1)e^{-2r}}{2\eta},~~c=\frac{2b-e^{-2r}}{2\sqrt{\eta}%
},\label{eq1}\\
b  &  =\frac{-\left\vert \eta-1\right\vert +\eta e^{2r}+e^{-2r}}%
{2[-e^{2r}\left\vert \eta-1\right\vert +\eta+1]}, \label{eq2}%
\end{align}
and the entanglement parameter $r\geq0$ is connected to the channel parameter
via the relation~\cite{Notation}
\begin{equation}
\nu=\frac{e^{-2r}}{2}(\eta+1). \label{eq3}%
\end{equation}

\section{Finite-resource teleportation stretching of an adaptive
protocol\label{sec2}}

Here we plug the previous finite-resource simulation into the tool of
teleportation stretching. We start by providing some necessary definitions on
adaptive protocols and secret-key capacity. Then, we review a general upper
bound (weak converse) based on the REE. Finally, following the recipe of
Ref.~\cite{PLOB,Ricc} we show how to use the finite-resource simulation to
simplify an adaptive protocol and reduce the REE bound to a single-letter quantity.

\subsection{Adaptive protocols and secret-key capacity}

The most general protocol for key generation is based on adaptive LOCCs, i.e.,
local operations assisted by unlimited and two-way classical communication.
Each transmission through the quantum channel is interleaved by two of such
LOCCs. The general formalism can be found in Ref.~\cite{PLOB} and goes as
follows. Assume that two remote users, Alice and Bob, have two local registers
of quantum systems (modes), $\mathbf{a}$ and $\mathbf{b}$, which are in some
fundamental state $\rho_{\mathbf{a}}\otimes\rho_{\mathbf{b}}$. The two parties
applies an adaptive LOCC $\Lambda_{0}$ before the first transmission.

In the first use of the channel, Alice picks a mode $a_{1}$ from her register
$\mathbf{a}$ and sends it through the channel $\mathcal{E}$. Bob gets the
output mode $b_{1}$ which is included in his local register $\mathbf{b}$. The
parties apply another adaptive LOCC $\Lambda_{1}$. Then, there is the second
transmission and so on. After $n$ uses, we have a sequence of LOCCs
$\{\Lambda_{0},\Lambda_{1},\ldots,\Lambda_{n}\}$ characterizing the protocol
$\mathcal{L}$\ and an output state $\rho_{\mathbf{ab}}^{n}$ which is
$\varepsilon$-close to a target private state~\cite{KD} with $nR_{n}$ bits.
Taking the limit of large $n$ and optimizing over the protocols, we define the
secret-key capacity of the channel
\begin{equation}
K(\mathcal{E})=\sup_{\mathcal{L}}\lim_{n}R_{n}~.
\end{equation}

\subsection{General upper bound}

According to Theorem $1$ (weak converse) in Ref.~\cite{PLOB}, a general upper
bound for $K(\mathcal{E})$ is given in terms of the REE of the output state
$\rho_{\mathbf{ab}}^{n}$
\begin{equation}
K(\mathcal{E})\leq E_{R}^{\star}(\mathcal{E}):=\sup_{\mathcal{L}}\lim_{n}%
\frac{E_{R}(\rho_{\mathbf{ab}}^{n})}{n}~. \label{REEbound}%
\end{equation}
Recall that the REE of a state $\rho$ is defined as $E_{R}(\rho)=\inf
_{\sigma_{\text{\textrm{sep}}}}S(\rho||\sigma_{\text{\textrm{sep}}})$, where
$\sigma_{\text{\textrm{sep}}}$ is a separable state and the relative entropy
is defined by $S(\rho||\sigma_{\text{\textrm{sep}}}):=\mathrm{Tr}[\rho
(\log_{2}\rho-\log_{2}\sigma_{\text{\textrm{sep}}})]$. These definitions can
be easily adapted for asymptotic states of bosonic systems.

Note that the first and simplest proof of Eq.~(\ref{REEbound}) can be found in
Ref.~\cite{PLOBv2} (the second arxiv version of Ref.~\cite{PLOB}). To avoid
potential misunderstandings or misinterpretations of this proof, we report
here the main points. For any protocol whose output $\rho_{\mathbf{ab}}^{n}%
$\ is $\varepsilon$-close (in trace norm) to target private state with rate
$R_{n}$ and dimension $d$, we may write%
\begin{equation}
nR_{n}\leq E_{R}(\rho_{\mathbf{ab}}^{n})+4\varepsilon\log_{2}d+2H_{2}%
(\varepsilon), \label{main}%
\end{equation}
where $H_{2}$ is the binary Shannon entropy. For distribution through a
discrete variable (DV) channel, whose output is a DV state, we may write
\begin{equation}
\log_{2}d\leq\alpha nR_{n}, \label{shieldUSED}%
\end{equation}
for some constant $\alpha$ [see also Eq.~(21) of Ref.~\cite{PLOBv2}]. The
exponential scaling in Eq.~(\ref{shieldUSED})\ comes from previous results in
Refs.~\cite{Matthias1a,Matthias2a}. The latter showed that, for any adaptive
protocol with rate $R_{n}$, there is another protocol with the same asymptotic
rate while having an exponential scaling for $d$.

The extension to a CV channel is achieved by a standard argument of truncation
of the output Hilbert space. After the last LOCC $\Lambda_{n}$, Alice and Bob
apply a truncation LOCC $\mathbb{T}_{d}$ which maps the output state
$\rho_{\mathbf{ab}}^{n}$ into a truncated\ version $\rho_{\mathbf{ab}}%
^{n,d}=\mathbb{T}_{d}(\rho_{\mathbf{ab}}^{n})$ with total dimension $d$. The
total protocol $\mathbb{T}_{d}\circ\mathcal{L}=\{\Lambda_{0},\Lambda
_{1},\cdots,\Lambda_{n},\mathbb{T}_{d}\}$ generates an output that is
$\varepsilon$-close to a DV private state with $nR_{n,d}$ bits. Therefore, we
may directly re-write Eq.~(\ref{main}) as%
\begin{equation}
nR_{n,d}\leq E_{R}(\rho_{\mathbf{ab}}^{n,d})+4\varepsilon\log_{2}%
d+2H_{2}(\varepsilon). \label{eeee}%
\end{equation}
Both the output and the target are DV states, so that we may again write
Eq.~(\ref{shieldUSED})~\cite{Notaprova}. Because $\mathbb{T}_{d}$ is a
trace-preserving LOCC, we exploit the monotonicity of the REE $E_{R}%
(\rho_{\mathbf{ab}}^{n,d})\leq E_{R}(\rho_{\mathbf{ab}}^{n})$ and rewrite
Eq.~(\ref{eeee}) as%
\begin{equation}
R_{n,d}\leq\frac{E_{R}(\rho_{\mathbf{ab}}^{n})+2H_{2}(\varepsilon
)}{n(1-4\alpha\varepsilon)}~.
\end{equation}
Taking the limit for large $n$ and small $\varepsilon$ (weak converse),
this\ leads to
\begin{equation}
\lim_{n}R_{n,d}\leq\lim_{n}n^{-1}E_{R}(\rho_{\mathbf{ab}}^{n}).
\end{equation}
The crucial observation is that in the right-hand side of the latter
expression, there is no longer dependence on the truncation $d$. Therefore, in
the optimization of $R_{n,d}$ over all protocols $\mathbb{T}_{d}%
\circ\mathcal{L}$ we can implicitly remove the truncation. Pedantically, we
may write%
\begin{align}
K(\mathcal{E})  &  =\sup_{d}\sup_{\mathbb{T}_{d}\circ\mathcal{L}}\lim
_{n}R_{n,d}\nonumber\\
&  \leq\sup_{\mathcal{L}}\lim_{n}n^{-1}E_{R}(\rho_{\mathbf{ab}}^{n}%
):=E_{R}^{\star}(\mathcal{E}). \label{main2}%
\end{align}

\begin{remark}
Note that the truncation argument was explicitly used in Ref.~\cite{PLOBv2} to
extend the bound to CV channels. See discussion after Eq.~(23) of
Ref.~\cite{PLOBv2}. There a cut-off was introduced for the total CV Hilbert
space at the output. Under this cutoff, the derivation for DV\ systems was
repeated, finding an upper bound which does not depend on the truncated
dimension (this was done by using the monotonicity of the REE exactly as
here). The cutoff was then relaxed in the final expression as above. The
published version~\cite{PLOB} includes other equivalent proofs but they have
been just given for completeness.
\end{remark}

\subsection{Simplification via teleportation stretching}

One of the key insights of Ref.~\cite{PLOB} has been the simplification of the
general bound in Eq.~(\ref{REEbound}) to a single-letter quantity. For bosonic
Gaussian channels, this was achieved by using teleportation stretching with
asymptotic simulations, where a channel is reproduced by CV teleportation over
a sequence of Choi-approximating resource states. Here we repeat the procedure
but we adopt the finite-resource simulation of Ref.~\cite{GerLimited}. Recall
that, differently from previous
approaches~\cite{B2main,Niset,MHthesis,Wolfnotes}, teleportation stretching
does not reduce a protocol into entanglement distillation but maintains the
task of the original protocol, so that adaptive key generation is reduced to
block (non-adaptive) key generation. See Ref.~\cite{TQCreview} for comparisons
and clarifications.

Assume that the adaptive protocol is performed over a phase-insensitive
Gaussian channel $\mathcal{E}_{\eta,\nu}$, so that we may use the simulation
in Eq.~(\ref{Finn}), where $\mathcal{T}_{\eta}$ is the Braunstein-Kimble
protocol with gain $\sqrt{\eta}$ and $\sigma_{\nu}$ is a zero-mean two-mode
Gaussian state, specified by Eqs.~(\ref{resState})-(\ref{eq3}). We may
re-organize an adaptive protocol in such a way that each transmission through
$\mathcal{E}_{\eta,\nu}$ is replaced by its resource state $\sigma_{\nu}$. At
the same time, each teleportation-LOCC $\mathcal{T}_{\eta}$ is included in the
adaptive LOCCs of the protocol, which are all collapsed into a single LOCC
$\bar{\Lambda}_{\eta}$ (trace-preserving after averaging over all
measurements). In this way, we may decompose the output state $\rho
_{\mathbf{ab}}^{n}:=\rho_{\mathbf{ab}}(\mathcal{E}_{\eta,\nu}^{\otimes n})$
as
\begin{equation}
\rho_{\mathbf{ab}}^{n}=\bar{\Lambda}_{\eta}(\sigma_{\nu}^{\otimes n})~.
\label{ggg}%
\end{equation}

The computation of $E_{R}(\rho_{\mathbf{ab}}^{n})$ can now be remarkably
simplified. In fact, we may write%
\begin{align}
E_{R}(\rho_{\mathbf{ab}}^{n})  &  =\inf_{\sigma_{\text{\textrm{sep}}}}%
S(\rho_{\mathbf{ab}}^{n}||\sigma_{\text{\textrm{sep}}})\nonumber\\
&  \overset{(1)}{\leq}\inf_{\sigma_{\text{\textrm{sep}}}}S[\bar{\Lambda}%
_{\eta}(\sigma_{\nu}^{\otimes n})||\bar{\Lambda}_{\eta}(\sigma
_{\text{\textrm{sep}}})]\nonumber\\
&  \overset{(2)}{\leq}\inf_{\sigma_{\text{\textrm{sep}}}}S(\sigma_{\nu
}^{\otimes n}||\sigma_{\text{\textrm{sep}}})=E_{R}(\sigma_{\nu}^{\otimes n}),
\end{align}
where: $(1)$ we consider the fact that $\bar{\Lambda}_{\eta}(\sigma
_{\text{\textrm{sep}}})$ form a subset of specific separable states, and $(2)$
we use the monotonicity of the relative entropy under the trace-preserving
LOCC $\bar{\Lambda}_{\eta}$. Therefore, by replacing in Eq.~(\ref{REEbound}),
we get rid of the optimization over the protocol (disappearing with
$\bar{\Lambda}_{\eta}$) and we may write
\begin{equation}
K(\mathcal{E}_{\eta,\nu})\leq\lim_{n}\frac{E_{R}(\sigma_{\nu}^{\otimes n})}%
{n}:=E_{R}^{\infty}(\sigma_{\nu})\leq E_{R}(\sigma_{\nu})~, \label{hhh}%
\end{equation}
where we use the fact that the regularized REE is less than or equal to the
REE. Thus, we may write the following theorem:

\begin{theorem}
Consider a phase-insensitive bosonic Gaussian channel $\mathcal{E}_{\eta,\nu}%
$, which is stretchable into a two-mode Gaussian state $\sigma_{\nu}$ as given
in Eqs.~(\ref{resState})-(\ref{eq3}). Its secret-key capacity must satisfy the
bound%
\begin{equation}
K(\mathcal{E}_{\eta,\nu})\leq E_{R}(\sigma_{\nu}):=\inf_{\sigma
_{\text{\textrm{sep}}}}S(\sigma_{\nu}||\sigma_{\text{\textrm{sep}}})~.
\label{plpl}%
\end{equation}

\end{theorem}

Note that the new bound in Eq.~(\ref{plpl}) cannot beat the asymptotic bound
established by Ref.~\cite{PLOB} for bosonic channels, i.e.,
\begin{equation}
K(\mathcal{E}_{\eta,\nu})\leq\inf_{\sigma_{\text{\textrm{sep}}}^{\mu}%
}\underset{\mu\rightarrow+\infty}{\lim\inf}S(\rho{_{\mathcal{E}_{\eta,\nu}%
}^{\mu}}||\sigma_{\text{\textrm{sep}}}^{\mu}), \label{REE_weaker}%
\end{equation}
where $\rho{_{\mathcal{E}_{\eta,\nu}}^{\mu}}$ is a Choi-approximating sequence
as in Eq.~(\ref{seqq}), and $\sigma_{\text{\textrm{sep}}}^{\mu}$ is an
arbitrary sequence of separable states converging in trace norm. This can be
seen from a quite simple argument~\cite{Andrea}. In fact, according to
Eqs.~(\ref{seqq}) and~(\ref{Finn}), we may write%
\begin{align}
\rho{_{\mathcal{E}_{\eta,\nu}}^{\mu}}  &  =\mathcal{I}\otimes\mathcal{E}%
_{\eta,\nu}(\Phi^{\mu})\nonumber\\
&  =\mathcal{I}\otimes\mathcal{T}_{\eta}(\Phi^{\mu}\otimes\sigma_{\nu}%
)=\Delta(\sigma_{\nu}),
\end{align}
where $\Delta$ is a trace-preserving LOCC. Therefore, $E_{R}(\rho
{_{\mathcal{E}_{\eta,\nu}}^{\mu}})\leq E_{R}(\sigma_{\nu})$ and this relation
is inherited by the bounds above. Notwithstanding this \textit{no go} for the
finite-resource simulation, we show that its performance is good and
reasonably approximates the infinite-energy bounds that are found via
Eq.~(\ref{REE_weaker}).

\section{Finite-resource bounds for phase insensitive Gaussian
channels\label{sec3}}

We now proceed by computing the REE in Eq.~(\ref{plpl}) for the class of
single-mode phase-insensitive Gaussian channels. For this, we exploit the
closed formula for the quantum relative entropy between Gaussian states which
has been derived in Ref.~\cite{PLOB} by using the Gibbs representation for
Gaussian states~\cite{Banchi}. Given two Gaussian states $\rho_{1}(u_{1}%
,V_{1})$ and $\rho_{2}(u_{2},V_{2})$, with respective statistical moments
$u_{i}$ and $V_{i}$, their relative entropy is
\begin{equation}
S(\rho_{1}||\rho_{2})=-\Sigma(V_{1},V_{1})+\Sigma(V_{1},V_{2})~,
\end{equation}
where we have defined
\begin{equation}
\Sigma(V_{1},V_{2}):=\frac{\ln\det\left(  V_{2}+\frac{i\Omega}{2}\right)
+\Tr(V_{1}G_{2})+\delta^{T}G_{2}\delta}{2\ln2}%
\end{equation}
with $\delta=u_{1}-u_{2}$ and $G_{2}=2i\Omega\coth^{-1}(2iV_{2}\Omega
)$~\cite{Banchi}, where the matrix $\Omega$ is the symplectic form.

The computation of the REE involves an optimization over the set of separable
states. Following the recipe of Ref.~\cite{PLOB} we may construct a good
candidate directly starting from the CM in Eq.~(\ref{resState}). This
separable state has CM with the same diagonal blocks as in Eq.~(\ref{resState}%
), but where the off-diagonal term is replaced as follows
\begin{equation}
c\rightarrow c_{\text{sep}}:=\sqrt{(a-1/2)(b-1/2)}~.
\end{equation}
By using this separable state $\tilde{\sigma}_{\text{sep}}$ we may write the
further upper bound
\begin{equation}
E_{R}(\sigma_{\nu})\leq\Psi(\mathcal{E}):=S(\sigma_{\nu}||\tilde{\sigma
}_{\text{sep}}).
\end{equation}
In the following, we compute this bound for the various types of
phase-insensitive Gaussian channels.

\subsection{{Thermal-loss channel}}

This channel can be modelled as a beam splitter of transmissivity $\eta$ where
the input signals are combined with a thermal environment such that the
quadratures transform according to $\hat{\mathbf{x}}\rightarrow\sqrt{\eta}%
\hat{\mathbf{x}}+\sqrt{1-\eta}\hat{\mathbf{x}}_{th}$, where $\hat{\mathbf{x}%
}_{th}$ is in a thermal state with $\bar{n}$ photons. In terms of the
statistical moments, the action of the thermal-loss channel $\mathcal{E}%
_{\eta,\bar{n}}$ can be described by the matrices in Eq.~(\ref{1mode}) with
parameter $\nu=(1-\eta)(\bar{n}+1/2)$. This means that the squeezing parameter
$r$ of the resource state now reads
\begin{equation}
r=\frac{1}{2}\ln\left[  \frac{\eta+1}{\left(  2\bar{n}+1\right)  (1-\eta
)}\right]  ~.
\end{equation}

\begin{figure}[ptbh]
\vspace{-0.0cm}
\par
\begin{center}
\includegraphics[width=0.4\textwidth]{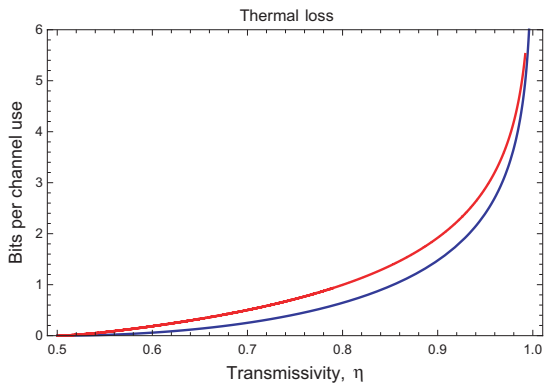} \vspace{-0.2cm}
\end{center}
\caption{Finite-resource bound $\Psi(\mathcal{E}_{\eta,\bar{n}})$ on the
secret-key capacity of the thermal loss channel (red upper curve) as a
function of the transmissivity $\eta$, compared with the infinite-energy bound
$\Phi(\mathcal{E}_{\eta,\bar{n}})$ (blue lower curve) derived in
Ref.~\cite{PLOB}. The curves are plotted for $\bar{n}=1$ thermal photons.}%
\label{Th}%
\end{figure}

By combining this relation with the ones in Eq.~(\ref{eq2}) and computing the
relative entropy, we find~\cite{analy} the finite-resource bound
$\Psi(\mathcal{E}_{\eta,\bar{n}})$ which is plotted in Fig.~\ref{Th} and
therein compared with the infinite-energy bound $\Phi(\mathcal{E}_{\eta
,\bar{n}})$ derived in Ref.~\cite{PLOB}. The latter is given by~\cite{PLOB}
\begin{equation}
\Phi(\mathcal{E}_{\eta,\bar{n}})=-\log_{2}[(1-\eta)\eta^{\bar{n}}]-h(\bar{n}),
\end{equation}
for $\bar{n}<\eta/(1-\eta)$ and zero otherwise, and we set $h(x):=(x+1)\log
_{2}(x+1)-x\log_{2}x$. It is clear that we have
\begin{equation}
K(\mathcal{E}_{\eta,\bar{n}})\leq\Phi(\mathcal{E}_{\eta,\bar{n}})\leq
\Psi(\mathcal{E}_{\eta,\bar{n}}),
\end{equation}
but the two upper bounds are reasonably close. 

\subsection{Noisy {amplifier channel}}

\begin{figure}[ptbh]
\begin{center}
\vspace{+0.0cm}
\includegraphics[width=0.43\textwidth]{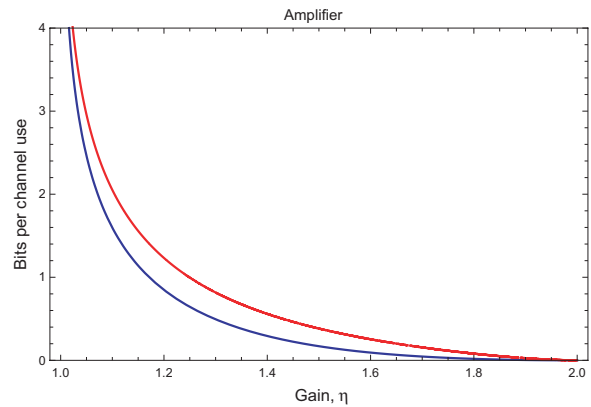} \vspace{-0.5cm}
\end{center}
\caption{Finite-resource bound $\Psi(\mathcal{E}_{\eta,\bar{n}})$ on the
secret-key capacity of the noisy amplifier channel (red upper curve) as a
function of the gain $\eta$, compared with the optimal bound for infinite
energy $\Phi(\mathcal{E}_{\eta,\bar{n}})$ (blue lower curve). The two curves
are plotted for $\bar{n}=1$ thermal photons.}%
\label{Ampli}%
\end{figure}

A noisy quantum amplifier is described by $\hat{\mathbf{x}}\rightarrow
\sqrt{\eta}\hat{\mathbf{x}}+\sqrt{\eta-1}\hat{\mathbf{x}}_{th}$, where
$\eta>1$ is the gain and $\hat{\mathbf{x}}_{th}$ is in a thermal state with
$\bar{n}$ photons. This channel $\mathcal{E}_{\eta,\bar{n}}$ is described by
the matrices in Eq.~(\ref{1mode}) with parameter $\nu=(\eta-1)(\bar{n}+1/2)$.
By repeating the previous calculations, we find~\cite{analy} the
finite-resource bound $\Psi(\mathcal{E}_{\eta,\bar{n}})$ plotted in
Fig.~\ref{Ampli} and where it is compared with the infinite-energy
bound~\cite{PLOB}%
\begin{equation}
\Phi(\mathcal{E}_{\eta,\bar{n}})=\log_{2}\left(  \frac{\eta^{\bar{n}+1}}%
{\eta-1}\right)  -h(\bar{n}),
\end{equation}
for $\bar{n}<(\eta-1)^{-1}$ and zero otherwise.

\subsection{{Additive-noise Gaussian channel}}

\begin{figure}[ptbh]
\begin{center}
\vspace{+0.1cm}
\includegraphics[width=0.42\textwidth]{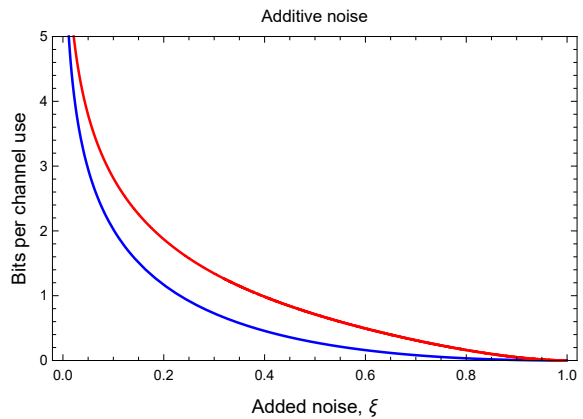} \vspace{-0.4cm}
\end{center}
\caption{Finite-resource bound $\Psi(\mathcal{E}_{\xi})$ on the secret-key
capacity of the additive noise Gaussian channel (red upper curve) as a
function of the added noise $\xi$, compared with the optimal bound for
infinite energy $\Phi(\mathcal{E}_{\xi})$ (blue lower curve).}%
\label{AddNoise}%
\end{figure}

Another important channel is represented by the additive-noise Gaussian
channel, which is the simplest model of bosonic decoherence. In terms of the
input-output transformations, the quadratures transforms according to
$\hat{\mathbf{x}}\rightarrow\hat{\mathbf{x}}+(z,z)^{T}$ where $z$ is a
classical Gaussian variable with zero mean and variance $\xi\geq0$. This
channel $\mathcal{E}_{\xi}$ is described by the matrices in Eq.~(\ref{1mode})
with $\eta=1$ and $\nu=\xi$. The finite-resource bound~\cite{analy}
$\Psi(\mathcal{E}_{\xi})$ on the secret key capacity is plotted in
Fig.~\ref{AddNoise} and compared with the infinite-energy bound~\cite{PLOB}
\begin{equation}
\Phi(\mathcal{E}_{\xi})=\frac{\xi-1}{\ln2}-\log_{2}\xi,
\end{equation}
for $\xi<1$, while zero otherwise. 

\subsection{{Pure-loss channel}}

For the pure-loss channel, the upper bound derived in the limit of infinite
energy~\cite{PLOB} coincides with the lower bound computed with the reverse
coherent information~\cite{RevCohINFO,ReverseCAP}. This means that we are able
to fully characterize the secret-key capacity for this specific bosonic
channel. This is also known as the Pirandola-Laurenza-Ottaviani-Banchi (PLOB)
bound~\cite{PLOB}
\begin{equation}
\mathcal{K}(\eta)=-\log_{2}(1-\eta)\simeq1.44\eta\text{ for }\eta\simeq0~,
\end{equation}
and fully characterizes the fundamental rate-loss scaling of point-to-point
quantum optical communications.

\begin{figure}[h]
\begin{center}
\vspace{-0.0cm}
\includegraphics[width=0.42\textwidth]{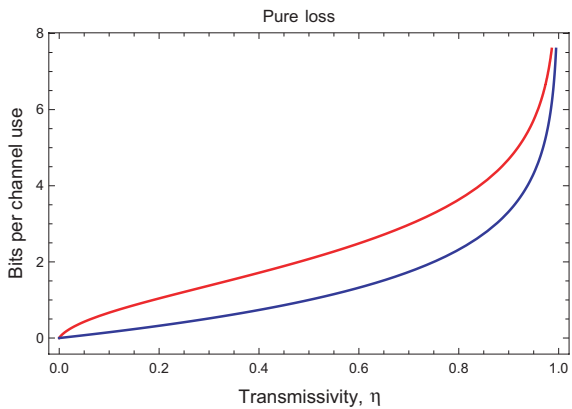} \vspace{-0.45cm}
\end{center}
\caption{Finite-resource bound $\Psi(\mathcal{E}_{\eta})$ on the secret-key
capacity of the pure-loss channel (red upper curve) as a function of the
transmissivity $\eta$, compared with its secret key capacity or PLOB bound
$K(\eta)=-\log_{2}(1-\eta)$ (blue lower curve).}%
\label{pureloss}%
\end{figure}

Consider now the finite-resource teleportation simulation of a pure-loss
channel. It is easy to check that we cannot use the parametrization in
Eq.~(\ref{eq2}). In fact, for a pure-loss channel, we have $\nu=(1-\eta)/2$ so
that Eq.~(\ref{eq3}) provides $e^{2r}=(1+\eta)/(1-\eta)$. Replacing the latter
in Eq.~(\ref{eq2}), we easily see that we have divergences (e.g., the
denominator of $b$ becomes zero). For the pure loss channel, we therefore use
a different simulation, where the resource state is a two-mode squeezed state
with CM~\cite{Andrea}%
\begin{equation}
\sigma_{\eta}=\left(
\begin{array}
[c]{cc}%
a\mathbf{I} & \sqrt{a^{2}-1/4}\mathbf{Z}\\
\sqrt{a^{2}-1/4}\mathbf{Z} & a\mathbf{I}%
\end{array}
\right)  ,~a=\frac{\eta+1}{2(1-\eta)}~.
\end{equation}
By exploiting this resource state, we derive~\cite{analy} the bound
$\Psi(\mathcal{E}_{\eta})$ shown in Fig.~\ref{pureloss}, where it is compared
with the secret-key capacity $K(\eta)$.

\section{Extension to repeater-assisted private communication\label{sec4}}

Here we extend the previous treatment to repeater-assisted private
communication. We consider the basic scenario where Alice $\mathbf{a}$ and Bob
$\mathbf{b}$\ are connected by a chain of $N$ quantum repeaters $\{\mathbf{r}%
_{1},\ldots,\mathbf{r}_{N}\}$, so that there are a total of $N+1$ quantum
channels $\{\mathcal{E}_{i}\}$ between them. Assume that these are
phase-insensitive Gaussian channels $\mathcal{E}_{i}:=\mathcal{E}_{\eta
_{i},\nu_{i}}$ with parameters $(\eta_{i},\nu_{i})$. The most general adaptive
protocol for key distribution through the chain is described in
Ref.~\cite{networkPIRS}\ and goes as follows.

Alice, Bob and all the repeaters prepare their local registers $\{\mathbf{a}%
,\mathbf{r}_{1},\ldots,\mathbf{r}_{N},\mathbf{b}\}$ into a global initial
state $\rho^{0}$ by means of a network LOCC $\Lambda_{0}$, where each node in
the chain applies LOs assisted by unlimited and two-way CCs with all the other
nodes. In the first transmission, Alice picks a system $a_{1}\in\mathbf{a}$
and sends it to the first repeater; after another network LOCC $\Lambda_{1}$,
the first repeater communicates with the second repeater; then there is
another network LOCC $\Lambda_{2}$ and so on, until Bob is eventually reached,
which terminates the first use of the chain.

After $n$ uses of the chain, we have a sequence of network LOCCs $\mathcal{L}$
defining the protocol and an output state $\rho_{\mathbf{ab}}^{n}$ for Alice
and Bob which approximates some target private state with $nR_{n}$ bits. By
taking the limit for large $n$ and optimizing over the protocols, we define
the end-to-end or repeater-assisted secret-key capacity~\cite{networkPIRS}%
\begin{equation}
K(\{\mathcal{E}_{i}\})=\sup_{\mathcal{L}}\lim_{n}R_{n}~.
\end{equation}
As shown in Ref.~\cite{networkPIRS}, we may extend the upper bound of
Eq.~(\ref{REEbound}). Then, we may use teleportation stretching and optimize
over cuts of the chain, to simplify the bound to a single-letter quantity.

\begin{figure}[ptbh]
\vspace{+0.1cm}
\par
\begin{center}
\includegraphics[width=0.42\textwidth]{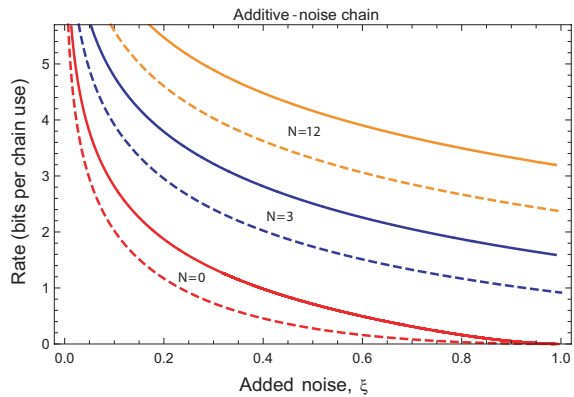} \vspace{-0.5cm}
\vspace{-1cm}\vspace{1cm}
\end{center}
\caption{Secret-key capacity of a chain of $N$ equidistant repeaters creating
$N+1$ additive-noise Gaussian channels with variances $\xi_{i}=\xi/(N+1)$.\ We
compare the finite-resource upper bound $\Psi(\{\mathcal{E}_{i}\})$ (solid
lines) with the infinite-energy upper bound $\Phi(\{\mathcal{E}_{i}%
\})$\ (dashed lines) for different values of $N$ as a function of the overall
added noise of the chain $\xi$.}%
\label{lklk}%
\end{figure}

The network-reduction technique of Ref.~\cite{networkPIRS} can be implemented
by using the specific finite-resource simulation of Eq.~(\ref{Finn}), which
leads to the following possible decompositions of the output state%
\begin{equation}
\rho_{\mathbf{ab}}^{n}=\bar{\Lambda}_{i}(\sigma_{\nu_{i}}^{\otimes
n}),~~\text{for any }i=1,\ldots,N,
\end{equation}
where $\bar{\Lambda}_{i}$ is a trace-preserving LOCC and\ $\sigma_{\nu_{i}}$
is the resource state associated with the $i$th Gaussian channel.\ By
repeating the derivation of Ref.~\cite{networkPIRS}, this leads to%
\begin{equation}
K(\{\mathcal{E}_{i}\})\leq\min_{i}E_{R}(\sigma_{\nu_{i}})\leq\min_{i}%
S(\sigma_{\nu_{i}}||\tilde{\sigma}_{i,\text{sep}}):=\Psi(\{\mathcal{E}%
_{i}\})~, \label{rete}%
\end{equation}
where $\Psi$ is the upper bound coming from our choice of the separable state
$\tilde{\sigma}_{i,\text{sep}}$ in the REE. This upper bound needs to be
compared with the one $\Phi(\{\mathcal{E}_{i}\})$ obtained in the limit of
infinite energy~\cite{networkPIRS}. As an example, consider an additive-noise
Gaussian channel with noise variance $\xi$. Let us split the communication
line by using $N$ \textquotedblleft equidistant\textquotedblright\ repeaters,
in such a way that each link is an additive-noise Gaussian channel
$\mathcal{E}_{i}$ with the same variance $\xi_{i}=\xi/(N+1)$. It is easy to
check that this is the optimal configuration for the repeaters. From
Eq.~(\ref{rete}), we derive $\Psi(\{\mathcal{E}_{i}\})=\Psi(\mathcal{E}%
_{\xi/(N+1)})$. This bound is plotted in Fig.~\ref{lklk} where we can se an
acceptable approximation of the corresponding infinite-energy bound
$\Phi(\{\mathcal{E}_{i}\})$. 

\section{Conclusion\label{sec5}}

\vspace{-0.1cm} In this work we have presented a design for the technique of
teleportation stretching~\cite{PLOB} for single-mode bosonic Gaussian
channels, where the core channel simulation~\cite{GerLimited} is based on a
finite-energy two-mode Gaussian state processed by the Braunstein-Kimble
protocol~\cite{teleCV} with suitable gains. Such an approach removes the need
for using an asymptotic simulation where the sequence of states approximates
the energy-unbounded Choi matrix of a Gaussian channel, even though the
infinite energy limit remains at the level of Alice's quantum measurement
which is ideally a CV\ Bell detection (i.e., a projection onto displaced EPR states).

Using this approach we compute the weak converse bound for the secret key
capacity of all phase-insensitive single-mode Gaussian channels, which include
the thermal-loss channel, the quantum amplifier and the additive-noise
Gaussian channel. We find that the bounds so derived are reasonably close to
the tightest known bound established in Ref.~\cite{PLOB} by using asymptotic
Choi matrices. We considered not only for point-to-point communication but
also a repeater-assisted scenario where Alice and Bob are connected by a chain
of quantum repeaters. The tools developed here may have other applications;
they may be applied to multi-point protocols~\cite{Multipoint} or to quantum
metrology~\cite{reviewMETRO}, e.g., to approximate the bounds for the adaptive
estimation of Gaussian channels established in Ref.~\cite{PirCo}.

\textit{Note added}.~Our work first appeared on the arXiv in June
2017~\cite{erxiv}. It has been revised after an imprecision in
Ref.~\cite{GerLimited}\ was fixed in Ref.~\cite{erratum}. Independently, a
related work~\cite{Kaur} directly built on the techniques of Ref.~\cite{PLOB},
but its claims are restricted to a point-to-point thermal-loss channel in the
non-asymptotic scenario.

\textit{Acknowledgments.}~This work has been supported by the EPSRC via the
`UK Quantum Communications Hub' (EP/M013472/1). The authors would like to
thank G. Adesso for comments on our first developments soon after the
appearance of Ref.~\cite{GerLimited}, and A. Mari for discussions on the
relations between the various bounds, and the finite-resource simulation of
the pure-loss channel. The authors also thank C. Lupo, G. Spedalieri, C.
Ottaviani, S. Tserkis, T. Ralph, and S. Lloyd.


\begin{thebibliography}{99}                                                                                               %


\bibitem {kimbleQnet}H. J. Kimble, Nature \textbf{453}, 1023 (2008).

\bibitem {pirsQnet}S. Pirandola and S. L. Braunstein, Nature \textbf{532}, 169 (2016).

\bibitem {Watrous}J. Watrous, \textit{The theory of quantum information}
(Cambridge University Press, Cambridge, 2018).


\bibitem {NiCh}M. A. Nielsen, and I. L. Chuang, \textit{Quantum computation
and quantum information} (Cambridge University Press, Cambridge, 2000).

\bibitem {QIbook}M. Hayashi, \textit{Quantum Information Theory: Mathematical
Foundation} (Springer-Verlag Berlin Heidelberg, 2017).

\bibitem {RMP}C. Weedbrook \textit{et al.,} Rev. Mod. Phys. \textbf{84}, 621 (2012).

\bibitem {SamRMPm}S. L. Braunstein, and P. Van Loock, Rev. Mod. Phys.
\textbf{77}, 513 (2005).

\bibitem {Alex}A. Serafini, F. Illuminati, and S. De Siena, Journal of Physics
B: Atomic, Molecular and Optical Physics \textbf{37}, L21 (2004).

\bibitem {Gerry}G. Adesso, S. Ragy, and A. R. Lee, Open Systems and
Information Dynamics \textbf{21}, 1440001 (2014).

\bibitem {BB84}C. H. Bennett, and G. Brassard, Proc. IEEE International Conf.
on Computers, Systems, and Signal Processing, Bangalore, pp. 175--179 (1984).

\bibitem {PLOB}S. Pirandola, R. Laurenza, C. Ottaviani, and L. Banchi, Nat.
Commun. \textbf{8}, 15043 (2017). See also arXiv:1510.08863 (2015).

\bibitem {Rep1}H.-J. Briegel, W. D\"{u}r, J. I. Cirac, and P. Zoller, Phys.
Rev. Lett. \textbf{81}, 5932-5935 (1998).

\bibitem {Rep2}W. D\"{u}r, H.-J. Briegel, J. I. Cirac, and P. Zoller, Phys.
Rev. A \textbf{59}, 169 (1999).

\bibitem {Rep3}L. M. Duan, M. D. Lukin, J. I. Cirac, and P. Zoller,Nature
(London) \textbf{414}, 413 (2001).

\bibitem {Rep4}Z. Zhao, T. Yang, Y.-A. Chen, A.-N. Zhang, and J.-W. Pan, Phys.
Rev. Lett. \textbf{90}, 207901 (2003).

\bibitem {Rep5}C. Simon, H. de Riedmatten, M. Afzelius, N. Sangouard, H.
Zbinden, and N. Gisin\textit{, }Phys. Rev. Lett. \textbf{98}, 190503 (2007).

\bibitem {Rep6}Z.-S. Yuan, Y.-A. Chen, B. Zhao, S. Chen, J. Schmiedmayer, and
J.-W. Pan, Nature \textbf{454}, 1098-1101 (2008).

\bibitem {Rep7}P. van Loock, N. L\"{u}tkenhaus, W. J. Munro, and K. Nemoto,
Phys. Rev. A \textbf{78}, 062319 (2008).

\bibitem {Rep8}R. Alleaume, F. Roueff, E. Diamanti, and N. L\"{u}tkenhaus, New
J. Phys. \textbf{11}, 075002 (2009).

\bibitem {Rep9}N. Sangouard, C. Simon, H. de Riedmatten, and N. Gisin, Rev.
Mod. Phys. \textbf{83}, 33 (2011).

\bibitem {Rep10}D. E. Bruschi, T. M. Barlow, M. Razavi, and A. Beige\textit{,
}Phys. Rev. A \textbf{90}, 032306 (2014).

\bibitem {Rep12}S. Muralidharan, J. Kim, N. L\"{u}tkenhaus, M. D. Lukin, and
L. Jiang, Phys. Rev. Lett. \textbf{112}, 250501 (2014).

\bibitem {Rep13}K. Azuma, K. Tamaki, and W. J. Munro, Nature Comm. \textbf{6},
10171 (2015).

\bibitem {Rep13bis}S. B\"{a}uml, M. Christandl, K. Horodecki, and A. Winter,
Nature Comm. \textbf{6}, 6908 (2015).

\bibitem {Rep14}D. Luong, L. Jiang, J. Kim, and N. L\"{u}tkenhaus\textit{,}
Appl. Phys. B \textbf{122}, 96 (2016).

\bibitem {Rep15}J. Dias and T. C. Ralph, Phys. Rev. A \textbf{95}, 022312 (2017).

\bibitem {Rep16}M. Pant, H. Krovi, D. Englund, and S. Guha, Phys. Rev. A
\textbf{95}, 012304 (2017).

\bibitem {Rep18}M. Christandl and A. Muller-Hermes, Communications in
Mathematical Physics \textbf{353}, 821-852 (2017).

\bibitem {bench5}F. Rozpedek, K. Goodenough, J. Ribeiro, N. Kalb, V. Caprara
Vivoli, A. Reiserer, R. Hanson, S. Wehner, and D. Elkouss, Quantum Sci.
Technol. \textbf{3}, 034002 (2018).

\bibitem {LoPiparo1}N. Lo Piparo, N. Sinclair, and M. Razavi, arXiv:1707.07814 (2017).

\bibitem {LoPiparo1bis}N. Lo Piparo, and M. Razavi, \textit{Memory-assisted
quantum key distribution immune to multiple-excitation effects}, Conference on
Lasers and Electro-Optics (CLEO), 5-10 June 2016.

\bibitem {LoPiparo2}N. Lo Piparo, M. Razavi, W. J. Munro, Phys. Rev. A
\textbf{96}, 052313 (2017).

\bibitem {MihirRouting}M. Pant, H. Krovi, D. Towsley, L. Tassiulas, L. Jiang,
P. Basu, D. Englund, and S. Guha, arXiv:1708.07142 (2017).



\bibitem {RevCohINFO}R. Garc\'{\i}a-Patr\'{o}n, S. Pirandola, S. Lloyd, and J.
H. Shapiro,\ Phys. Rev. Lett. \textbf{102}, 210501 (2009).

\bibitem {ReverseCAP}S. Pirandola, R. Garc\'{\i}a-Patr\'{o}n, S. L.
Braunstein, and S. Lloyd, Phys. Rev. Lett. \textbf{102}, 050503 (2009).

\bibitem {RMPrelent}V. Vedral, Rev. Mod. Phys.\textit{ }\textbf{74}, 197 (2002).

\bibitem {VedFORMm}V. Vedral, M. B. Plenio, M. A. Rippin, and P. L. Knight,
Phys. Rev. Lett. \textbf{78}, 2275-2279 (1997).

\bibitem {Pleniom}V. Vedral, and M. B. Plenio, Phys. Rev. A \textbf{57}, 1619 (1998).

\bibitem {Donaldmain}B. Synak-Radtke and M. Horodecki, J. Phys. A: Math. Gen.
\textbf{39}, L423-L437 (2006).

\bibitem {Matthias1a}M. Christandl, A. Ekert, M. Horodecki, P. Horodecki, J.
Oppenheim, and R. Renner, Lecture Notes in Computer Science \textbf{4392},
456-478 (2007). See also arXiv:quant-ph/0608199v3 for a more extended version.

\bibitem {Matthias2a}M. Christandl, N. Schuch, and A. Winter, Comm. Math.
Phys. \textbf{311}, 397-422 (2012).

\bibitem {B2main}C. H. Bennett, D. P. DiVincenzo, J. A. Smolin, and W. K.
Wootters, Phys. Rev. A \textbf{54}, 3824-3851 (1996).

\bibitem {Niset}J. Niset, J. Fiurasek, and N. J. Cerf, Phys. Rev. Lett.
\textbf{102}, 120501 (2009).

\bibitem {MHthesis}A. M\"{u}ller-Hermes, Master's thesis (Technical University
of Munich, 2012).

\bibitem {Wolfnotes}M. M. Wolf,\textit{Quantum Channels \& Operations},
available at https://www-m5.ma.tum.de/foswiki/pub/M5/ Allgemeines/MichaelWolf/QChannelLecture.pdf.

\bibitem {TQCreview}S. Pirandola, S. L. Braunstein, R. Laurenza, C. Ottaviani,
T. P. W. Cope, G. Spedalieri, and L. Banchi, Quantum Sci. Technol. \textbf{3},
035009 (2018).

\bibitem {PLB}S. Pirandola, R. Laurenza, and S. L. Braunstein,
\textit{Teleportation simulation of bosonic Gaussian channels: Strong and
uniform convergence}, arXiv:1712.01615 (2017).

\bibitem {tele}C. H. Bennett, G. Brassard, C. Crepeau, R. Jozsa, A. Peres, and
W. K. Wootters, Phys. Rev. Lett. \textbf{70}, 1895 (1993).

\bibitem {telereview}S. Pirandola \textit{et al.}, Nature Photon. \textbf{9},
641-652 (2015).

\bibitem {SougatoBowen}G. Bowen and S. Bose, Phys. Rev. Lett. \textbf{87},
267901 (2001).

\bibitem {nonPauli}T. P. W. Cope, L. Hetzel, L. Banchi, and S. Pirandola,
Phys. Rev. A \textbf{96}, 022323 (2017).

\bibitem {Leung}D. Leung and W. Matthews, IEEE Trans. Info. Theory
\textbf{61}, 4486-4499 (2015).

\bibitem {teleCV}S. L. Braunstein and H. J. Kimble, Phys. Rev. Lett.
\textbf{80}, 869--872 (1998).

\bibitem {CiracCV}G. Giedke and J. I. Cirac, Phys. Rev. A \textbf{66}, 032316 (2002).

\bibitem {GerLimited}P. Liuzzo-Scorpo, A. Mari, V. Giovannetti, and G. Adesso,
Phys. Rev. Lett. \textbf{119}, 120503 (2017).

\bibitem {NoteBELL}More generally, one also needs to consider sequences of
LOCCs $\mathcal{T}^{\mu}$, so that the asymptotic simulation reads
$\mathcal{E}(\rho)=\lim_{\mu}\mathcal{T}^{\mu}(\rho\otimes\rho_{\mathcal{E}%
}^{\mu})$. For simplicity we omit this technicality, referring the reader to
Ref.~\cite{PLOB} for more details.

\bibitem {NoteBELL2}Note that, in the simulation of Eq.~(\ref{Finn}), one uses
a Braunstein-Kimble protocol with an ideal CV Bell detection. The latter is an
asymptotic measurement defined in the limit of infinite squeezing, i.e.,
infinite energy. For this reason, the finite-energy aspect of the simulation
in Eq.~(\ref{Finn}) only refers to the resource state.

\bibitem {teleMANCIO}S. Pirandola, and S. Mancini, Laser Physics \textbf{16},
1418 (2006).

\bibitem {Notation}Note that, with respect to the fomulas of
Ref.~\cite{GerLimited},\ we have an extra $1/2$ factor in Eqs.~(\ref{resState}%
) and~(\ref{eq3}). This is due to the different notation we adopt here. We set
the quadrature variance of the vacuum state to be $1/2$, while it was equal to
$1$ in Ref.~\cite{GerLimited}.

\bibitem {Ricc}S. Pirandola and R. Laurenza, \textit{General Benchmarks for
Quantum Repeaters}, arXiv:1512.04945 (15 Dec 2015).

\bibitem {KD}K. Horodecki, M. Horodecki, P. Horodecki, and J. Oppenheim, Phys.
Rev. Lett. \textbf{94}, 160502 (2005).

\bibitem {PLOBv2}S. Pirandola, R. Laurenza, C. Ottaviani, and L. Banchi,
\textit{The Ultimate Rate of Quantum Communications},\ arXiv:1510.08863v2 (8
Dec 2015).

\bibitem {Notaprova}Since the Hilbert space is finite-dimensional, the proof
of Refs.~\cite{Matthias1a,Matthias2a} automatically applies, i.e., the
protocol can be stopped after $n_{0}$ uses, and then repeated $m$ times in an
i.i.d. fashion, with $n=n_{0}m$. Key distillation applied to the $m$ DV output
copies implies a number of bits of CCs which is linear in $m$ which, in turn,
leads to an exponential scaling of $d$ in $n$.

\bibitem {Andrea}Andrea Mari, private communication.

\bibitem {Banchi}L. Banchi, S. L. Braunstein, and S. Pirandola, Phys. Rev.
Lett. \textbf{115}, 260501 (2015).

\bibitem {analy}The analytical expression is too cumbersome to be reported here.

\bibitem {networkPIRS}S. Pirandola, S. \textit{Capacities of repeater-assisted
quantum communications}, arXiv:1601.00966 (5 Jan 2016).

\bibitem {equi}It is easy to check that this is the optimal configuration for
the repeaters.

\bibitem {Multipoint}R. Laurenza, and S. Pirandola, Phys. Rev. A \textbf{96},
032318 (2017).

\bibitem {reviewMETRO}R. Laurenza, C. Lupo, G. Spedalieri, S. L. Braunstein,
and S. Pirandola, Quantum Meas. Quantum Metrol. \textbf{5}, 1-12 (2018).

\bibitem {PirCo}S. Pirandola, and C. Lupo, Phys. Rev. Lett. \textbf{118},
100502 (2017).

\bibitem {erxiv}R. Laurenza, S. L. Braunstein, and S. Pirandola,
arXiv:1706.06065v1 (June 2017).

\bibitem {erratum}P. Liuzzo-Scorpo, A. Mari, V. Giovannetti, and G. Adesso,
Phys. Rev. Lett. \textbf{120}, 029904(E) (2018).

\bibitem {Kaur}E. Kaur, and M. M. Wilde, arXiv:1706.04590v2.
\end{thebibliography}
\end{document}